\documentclass[aps,prd,twocolumn,preprintnumbers,nofootinbib,longbibliography]{revtex4-2}
\usepackage[utf8]{inputenc}
\usepackage{uniinput}
\usepackage{amsmath,mathtools,mathrsfs,hyperref}
\hypersetup{
	colorlinks=true,
	allcolors=blue
}

\renewcommand{\d}{\mathrm{d}}

\newcommand{\eg}{\textit{e.g.}}
\newcommand{\bea}{\begin{eqnarray}}
	\newcommand{\eea}{\end{eqnarray}}
\newcommand{\be}{\begin{equation}}
	\newcommand{\ee}{\end{equation}}

\newcommand{\orderMksecond}{\mathscr{O}\left(k^{-2}M^{-2}\right)}

\newcommand{\arxivA}{\href{https://arxiv.org/abs/1705.10172}{arXiv:1705.10172}~}
\newcommand{\arxivB}{\href{https://arxiv.org/abs/1712.00511}{arXiv:1712.00511}}
\newcommand{\arxivC}{\href{https://arxiv.org/abs/2109.09814}{arXiv:2109.09814}~}

\begin{document}

\title{Re-examining the stability of rotating horizonless\\ black shells mimicking Kerr black holes}

\author{Ulf Danielsson~}
\email{ulf.danielsson@physics.uu.se}
\affiliation{Institutionen för fysik och astronomi,
	Uppsala Universitet, Box 803, SE-751 08 Uppsala, Sweden\\}

\author{Suvendu Giri~}
\email{suvendu.giri@unimib.it}
\affiliation{Dipartimento di Fisica, Università di Milano-Bicocca, I-20126 Milano, Italy}
\affiliation{INFN, sezione di Milano-Bicocca, I-20126 Milano, Italy}

\preprint{UUITP-49/21}

\begin{abstract}
\noindent
In \arxivA a string theory inspired alternative to gravitational collapse was proposed, consisting of a bubble of AdS space made up of ingredients from string theory. These ultra compact objects are $9/8$ times the size of the corresponding Schwarzschild black hole, but being within the photosphere are almost indistinguishable from them. Slowly rotating counterparts of these black shells were constructed in \arxivB, which closely mimic a Kerr black hole, but have a quadrupole moment that differs from Kerr.
Recently, \arxivC studied the dynamical stability of the stationary black shells against radial perturbations and accretion of matter, and examined a two parameter family of fluxes required for stability.
In this paper, we re-examine the rotating black shells with particular attention to the constraints imposed by this dynamical analysis for non-rotating shells. Extrapolating these results to rotating shells, we find that they can indeed support themselves at a critical point in the gravitational potential. Additionally, requiring that they settle back to their new Buchdahl radius after accreting matter, uniquely fixes the fluxes required for dynamical stability. The flux parameters turn out to have an extremely simple form, and fulfil one of the constraints for perturbative radial stability while exactly saturating the other. The preferred quadrupole moment that we find, given some physical assumptions, is $7\%$ less than Kerr.
\end{abstract}

\maketitle

\section{Introduction}\label{sec:intro}

\noindent
Black holes continue to challenge our \textsc{uv} understanding of physics. On one hand, the existence of an event horizon gives rise to the information paradox \cite{Hawking:1976ra} (see \cite{2017tasi.conf..353P} for a review), while on the other hand, the spacetime singularity at the center of the black hole has evaded a satisfactory understanding in our best developed theories of \textsc{uv} physics. The enormous entropy of black holes  seems to have an origin at the horizon \cite{Hawking:1976de}, but being featureless à la the no-hair theorem \cite{Ruffini:1971bza}, we do not have a complete understanding of its thermodynamic origin. While string theory has had success in accounting for the microscopic entropy of supersymmetric black holes \cite{Strominger:1996sh}, it is less clear what happens in the case of the non-supersymmetric black holes of the real world. 

Attempts to understand how information can be preserved even in the presence of a genuine horizon, such as the idea of black hole complementarity proposed in \cite{Susskind:1993if}, has run into problems. In fact, in \cite{Almheiri:2012rt}, it was argued that full compatibility with quantum mechanics requires the horizon to be replaced by a firewall. For other discussions about possible new physics see \eg \cite{Mathur:2009hf,tHooft:1991uqr,Lunin:2002qf,Giusto:2004id} and the review in \cite{Cardoso:2019rvt}. This is the motivation behind proposals replacing black holes with other objects that lack a horizon, and in this way evade the information paradox.

Usually, the surface of such an object hides very close to where the horizon would have been, making it very difficult to distinguish them from ordinary black holes. 
In \cite{Danielsson:2017riq}, however, we proposed a model where the space time is unstable against the formation of bubbles of AdS, which then replace Schwarzschild black holes. These bubbles  have a radius equal to $9R_s /8$, commonly known as the Buchdahl radius, which is macroscopically larger than the Schwarzschild radius. In our paper we showed how the necessary matter components to build such bubbles arise in a natural way from string theory.

In \cite{Danielsson:2017pvl} we generalized our model to include slow rotation, and proposed that Kerr black holes are replaced by rotating black shells. Since the black shells lack horizons the exterior metric need not be the one of Kerr. In fact, by considering metrics more general than Kerr we found a quadrupole moment that differed from the one of Kerr.\footnote{The improved analysis of the present paper picks a physically better motivated case, and gives a different prediction than the one of \cite{Danielsson:2017pvl}.}

An important part of our model is a source term that transfers energy between the different components, preserving the total energy. Without such a term the black shell cannot be stable. In \cite{Danielsson:2017riq} we argued for the existence of a suitable source term in the quasi-static case. In \cite{Danielsson:2021ykm} this analysis was reconsidered in the case of physical accretion of matter on to the black shell. Analytical constraints on the source term were obtained, and checked against numerical calculations. It was shown that the source term can indeed be tuned so that the black shell grows in radius and stabilizes at the new Buchdahl radius when matter is falling on to it.

In this paper we revisit our analysis of the rotating black shell confronting the constraints on the dynamics obtained in \cite{Danielsson:2021ykm}. It turns out that the combined analysis uniquely fixes the parameters determining the source. Furthermore, despite the complexity of the expressions involved in describing the rotating case, the final result is remarkably simple. The preferred value for the quadrupole moment that we find in our analysis is, given some physical assumptions, given by $-0.926 a²M$, and thus about $7\%$ less in magnitude than the one of Kerr.

The outline of the paper is as follows. In section \ref{sec:review} we review the non-rotating black shell as well as revisit the rotating one. In section \ref{sec:source} we discuss the role of the source term making use of the results in \cite{Danielsson:2021ykm}. Finally, we conclude in section \ref{sec:conclusions} with an outlook towards future work. 

\section{Black holes as black shells}\label{sec:review}

\subsection{General principles and the non-rotating case}\label{sec:nonrotating}

\noindent
The starting point is a metastable Minkowski spacetime that can decay non-perturbatively through tunneling, into a stable AdS vacuum. This takes place through the nucleation of a bubble containing the true vacuum, via a Coleman-de Luccia gravitational instanton \cite{Coleman:1980aw}. In the vacuum, the probability of such a nucleation event is exponentially suppressed. However, when matter collapses, threatening to form a black hole, such a phase transition can be triggered so that a bubble forms and captures the in-falling matter, which can attach itself to the bubble wall and turn into a gas of open strings. This increase in entropy can make the process inevitable rather than unlikely.

In \cite{Danielsson:2017riq} we showed how all required matter components can be obtained from string theory. The bubble wall itself is in the form of a brane, for instance a 5-brane wrapping an internal 3-cycle, which carries magnetic flux. Alternatively, this flux  can be thought of in terms of dissolved branes of a lower dimensionality, for instance 3-branes wrapping internal 3-cycles. Massless vibrations in the flux fields within the brane give rise to radiation on top of the shell. It is into this radiation that the in-falling matter converts when the bubble nucleates. The number of massless degrees of freedom is furthermore governed by the rank of the magnetic fluxes, i.e., number of dissolved branes. An important part of our model is that the radiation gas will be heated to the local Unruh temperature when the system is in equilibrium. 

With these components, we were able to solve the Israel-Lanczos-Sen junction conditions \cite{Israel:1966rt,Lanczos:1927,Sen:1924} and show that the black shell has a critical point at a radius equal to $9R_s /8$, commonly known as the Buchdahl radius (where $R_s \coloneqq 2M$ is the Schwarzschild radius). The junction conditions relate the energy density on the shell with the geometry in the bulk. The first of these ensures that the induced metric $(h_{ij})$ is well defined, while the second gives the energy momentum tensor on the shell $(S_{ij})$ in terms of its extrinsic curvature of the shell $(K_{ij})$ embedded in the bulk\footnote{Throughout this paper Greek indices $μ,ν$ refer to bulk coordinates, while Latin indices $i,j$ refer to those on the shell.}
\begin{equation}\label{eq:jc}
	Δh_{ij} = 0, \quad  S_{ij} = -\frac{1}{8π}\left( ΔK_{ij} - ΔK h_{ij} \right),
\end{equation}
where $Δ$ denotes the difference in the corresponding quantities across the shell, and $K \coloneqq K_{ij}h^{ij}$ is the trace of the extrinsic curvature.
These ensure that the composite spacetime which includes the inside, the outside and the surface of the bubble together solves Einstein's equations. For a bubble of radius $r=R$ with AdS space (with radius $1/k$) in its interior, and Schwarschild metric (with mass $M$) outside, the junction conditions give
\begin{equation}\label{eq:firstorderstressenergy}
	\begin{split}
		S^t{}_t &= -ρ = \frac{1}{4π R}\left(\sqrt{1-\frac{2M}{R}}-\sqrt{1+k² R²}\right),\\
		S^θ{}_θ = S^ϕ{}_ϕ &= p =  \frac{1}{8πR}\left(\frac{1-M/R}{\sqrt{1-2M/R}}-\frac{1+2k²R²}{\sqrt{1+k²R²}}\right),
	\end{split}
\end{equation}
where $ρ$ and $p$ denote the density and pressure respectively, of the perfect fluid constituting the shell.
The metric induced on the shell is 
\begin{equation}\label{eq:inducedmetric0}
	d s²_\textrm{ind} = -\left(1-\frac{2M}{R}\right) d t² + R² d θ² + R² \sin ²θ d ϕ².
\end{equation}
Based on physical arguments presented in \cite{Danielsson:2017riq}, we demand that the stress tensor can be split into the constituents above, namely a brane, gas and stiff matter with equations of state $p_τ=-ρ_τ, p_g=ρ_g/2$, and $p_s=ρ_s$ respectively. For the shell at the critical radius $R=9R_s/8$, this gives the individual components as:
\begin{equation}\label{eq:firstorderSsplit}
	\begin{split}
		\left(S_{\textrm{brane}}\right)^i _j &= \left(-\frac{k}{4π}+\frac{2}{27Mπ}-\frac{1}{81kM²π}\right)δ^i _j,\\
		\left(S_{\textrm{gas}}\right)^i _j &= \frac{1}{18Mπ}u^i u_j+\frac{1}{54Mπ}δ^i _j,\\
		\left(S_{\textrm{stiff}}\right)^i _j &= \frac{2}{81kM²π}u^i u_j+\frac{1}{81kM²π}δ^i _j,
	\end{split}
\end{equation}
where $u^i$ is the normalized velocity vector of the stationary shell $u^i = 3\left(1,0,0\right)$. Note that the metric used is the induced metric of eq (\ref{eq:inducedmetric0}) evaluated at $R=9R_s/8$. However, without any other ingredients in the model, this critical point turns out to be unstable and with the slightest perturbation, the shell will either start expanding or collapse into a black hole. We proposed in our paper a coupling between the brane tension and the gas such that energy can be transferred between the two, changing the tension and the density of the gas in response to a perturbation. 

This is accomplished through a source term in the continuity equation for the components that preserves the total energy. It was argued that a natural quantity that could govern such a source is the local Unruh temperature. Any deviation in the temperature of the system away from the Unruh temperature should initiate an energy transfer aimed at equilibrating the temperature. It was furthermore shown how such a process achieved stability in the case of quasi-static perturbations. In \cite{Danielsson:2017riq} we studied the quasi-static case and argued that one can find such a source term.

In \cite{Danielsson:2021ykm}, the stability of a non-rotating black shell was carefully studied, and it was shown that the quasi-static approach in \cite{Danielsson:2017riq} was not sufficient. The reason is corrections to the Unruh temperature due to the motion and acceleration of the shell. Any physical perturbation, such as the impact of matter, will induce motion that is not quasi-static in any limit. In \cite[eq (86)]{Danielsson:2021ykm}, the source term was therefore generalized to
\be\label{eq:flux_orig}
j \equiv 3\rho_g \left(\alpha\ \dot{a}_{_R}/a_{_{R}} + \beta\ F/2\right),
\ee
where $a_{_R}$ is the Unruh temperature, and the dot denotes the time derivative. We have defined $F=\dot{A}/A$, where $A$ is the area of the shell. $α$ and $β$ are constants that parametrize respectively, the response of the shell to a change in its Unruh temperature, and the response to a change in its area under compression or relaxation. The quasi-static case corresponds to $\alpha=\beta=1$. An analysis of linear perturbations proportional to $e^{i \omega \tau}$ with this source, led to a frequency given by \cite[eq (B.18)]{Danielsson:2021ykm}
\be \label{eq:pert}
\begin{split}
\omega \approx &\, \frac{128\pi\zeta_0}{27(4-9\alpha)m}\, \cdot \\
&\big(i \pm \sqrt{-1+9(4-9\alpha)(6\alpha - \beta-2)/(16\pi\zeta_0)^2}\big) ,
\end{split}
\ee
where we have introduced a non-zero viscosity given by $\zeta_0$, and $m$ is the mass of the bubble. We have taken the limit $m \gg 1/k$. From this one finds that stability with damping requires $\alpha < 4/9$ and $6\alpha - \beta > 2$. This is not all that is needed. Studying an impulsive accretion event, it was found that one must impose (in the large $m$ limit)
\be\label{eq:Buchdahl_constraint}
\alpha = \beta +2/3 .
\ee
This is to make sure that the shell, after accreting the in-falling matter, settles at the new Buchdahl radius. These constraints will play an important role for the rotating black shell. We will come back to this after reviewing the rotating case.

\subsection{Revisiting the rotating black shell}\label{sec:rotating}

\noindent
In \cite{Danielsson:2017pvl} we studied a slowly rotating black shell. To do this we considered solutions to the Einstein equations generalizing the Kerr-metric outside of the rotating shell, and generalizing the AdS-metric inside of the rotating shell. On the outside we add to the Kerr metric a superposition of a metric similar to the one of Novikov and Manko \cite{Manko_1992} (parametrized by $q$) and the one of Hartle and Thorne \cite{Hartle:1968si} (parametrized by $p$), given by:
\begin{equation}\label{eq:external}
	\begin{split}
		\d s² =& -g_{tt} \left(1+a²\left(q λ+ p χ\right)\right) \d t² \\
		&+ g_{rr} \left(1+a² \left(q ν- p χ\right)\right) \d r² \\
		& + g_{θθ} \left(1+a² \left(q ν+p ψ\right)\right) \d θ²\\
		&+ g_{ϕϕ} \left(1-a² \left(q λ-p ψ\right)\right) \d ϕ² \\
		&+ 2 g_{tϕ} \d t \d ϕ +  Δg_{μν}d x^μ d x^ν ,
	\end{split}
\end{equation}
where $g_{\mu \nu}$ is the Kerr metric with mass $M$ and spin $a$, while the perturbing functions $λ,ν,χ$ and $ψ$ are given by:
\begin{equation}\label{eq:external2}
	\begin{split}
		λ =& \left[(4M²-4Mr+2r^2)\mathscr{L}-4M(r-M)(3 \cos²θ-1) \right.\\
		&\left. -2\mathscr{L}(2M²-6Mr+3r²)\cos²θ \right]\frac{1}{M²} ,\\
		ν =& \left[ 2\mathscr{L}\cos²θ \left(6 M²-10Mr+3r²\right)+4M(5M-r) \right.\\
		&\left. -2\mathscr{L}\left(6 M²-6Mr+r²\right)+4M\cos²θ (3r-7M) \right]\frac{1}{M²} ,\\
		χ =& -\frac{5}{8 M²}\frac{\left(3\cos ²θ-1\right)}{r\left(r-2M\right)}\\
		&\left[2M\left(r-M\right)\left(2M²+6Mr-3r²\right)-3r²\left(r-2M\right)²\mathscr{L}\right],\\
		ψ =& \frac{5}{8 M²}\frac{\left(3\cos ²θ-1\right)}{r}\\
		&\left[2M\left(2M²-3Mr-3r²\right)-3r\left(r²-2M²\right)\mathscr{L}\right].
	\end{split}
\end{equation}
where $\mathscr{L} \coloneqq \log\left(1-2M/r\right)$.
Here we have also added an extra contribution, which solves Einstein's equations, to make sure that the mass of the system remains $M$, given by 
\begin{equation}
\begin{split}
	&Δg_{μν}d x^μ d x^ν = -\frac{16}{3}a²Mq\, \cdot \\
	& \left(\frac{r-M}{r²} d t²+\frac{r-M}{\left(r-2M\right)²} d r²  +r d θ²+r\sin ²θ d ϕ²\right).
\end{split}
\end{equation}
To have full control over the metric that we use, we have calculated its Geroch-Hansen multiple moments \cite{Geroch:1970cd,Hansen:1974zz}. The above metric has\footnote{Without the added piece, $\mathscr{M}_0$ would have had an extra contribution $(8/3)a²Mq$.}
\begin{equation}\label{eq:quadrupole}
	\begin{split}
	\mathscr{M}_0 &= M, \mathscr{J}_0 = 0,\\
	\mathscr{M}_1 &= 0, \mathscr{J}_1 = aM,\\
	\mathscr{M}_2 &= -a²M-\frac{2}{15}a²M³\left(16q-15p\right),
	\mathscr{J}_2 = 0, 
\end{split}
\end{equation}
with higher moments being of the form $\mathscr{M}_{2k} = \left(-1\right)^k M a^{2k}$, and $\mathscr{J}_{2k+1} = \left(-1\right)^k M a^{2k+1}$ (which come from the Kerr metric) plus terms containing $p$ and $q$ that appear with $a^n$ where $n>2$. Thus we see how $q$ and $p$ give rise to a deviation in the quadrupole moment from the one of Kerr. Our goal is to determine these parameters.

Inside the bubble, we construct a generalization of the AdS metric of the following form
\begin{equation}\label{eq:interiorold}
	\begin{split}
		\d s² =& -\left(1+k² r²\right) \left(1+a²\left(c₁μ₁+c₂σ₁\right)\right) \d t² \\
		&+ \left(1+k² r²\right)^{-1} \left(1+a²\left(c₁μ₂+c₂σ₂\right)\right) \d r²\\
		& + r² \left(1+a²\left(c₁μ₃+c₂σ₃\right)\right) \d θ² \\
		&+ r² \sin ²θ \left(1+a²\left(c₁μ₄+c₂σ₄\right)\right) \d ϕ²,
	\end{split}
\end{equation}
where the functions $μ_i, σ_i$ are given by
\begin{equation}\label{eq:insideHT}
	\begin{split}
		μ₁ &= f₁ \mathrm{P}_2\left(\cos θ\right),\, 
		μ₂ = -f₁ \mathrm{P}_2\left(\cos θ\right),\, \\
		μ₃ &= f₃ \mathrm{P}_2\left(\cos θ\right),\, 
		μ₄ = f₃ \mathrm{P}_2\left(\cos θ\right),\\
		σ₁ &= g₁ \left(3\cos ²θ -1\right),\, 
		σ₂ = \frac{r}{6} g₃^{\prime}\left(3\cos ²θ -1\right),\, \\
		σ₃ &= g₃ \cos ²θ,\, 
		σ₄ = g₃ \left(2\cos ²θ -1\right),
	\end{split}
\end{equation}
with $f_1,f_3,g_1$ and $g_3$ further given by
\begin{equation}
	\begin{split}
		f₁ &= \frac{5 k² r^2+3}{4 k^6 r^4+4 k^4 r^2} -\frac{3 \left(k² r^2+1\right) \textrm{arctan}(kr)}{4 k^5 r^3},\\
		f₃ &= \frac{4 k² r^2-3}{4 k^4 r^2} -\frac{3\left(k² r^2-1\right) \textrm{arctan}(kr)}{4 k^{5} r^3},\\
		g₁ &= -\frac{2 k^4 r^4-2 k² r^2-3}{6 k^6 r^4+6 k^4 r^2}-\frac{\textrm{arctan}(kr)}{2 k^5 r^3},\\
		g₃ &= \frac{k² r^2-3}{3 k^4 r^2}+\frac{\textrm{arctan}(kr)}{k^5 r^3}.
	\end{split}
\end{equation}
We then move on to solve the first junction condition to order $a^2$ in the spin for a shell that is located at coordinate radius $r_\textrm{out}$ from the outside, and $r_\textrm{in}$ from the inside
\begin{equation}
\begin{split}
	r_\textrm{out} &= R+Ra^2\left( m_1 + m_2 \cos 2 \theta \right),\\
	r_\textrm{in} &= R+Ra^2\left( n_1 + n_2 \cos 2 \theta \right).
\end{split}
\end{equation}
This relates parameters of the metric inside the bubble to those on the outside, in terms of those on the shell. In particular, it gives $c_1$ and $c_2$ in terms of $p, q$ and the difference $\left(m_1-n_1\right)$:
\begin{equation}
	\begin{split}
		c_1 =& \frac{k^{5/2}M}{96 \left[ 9 k M (27 k M^2 -4) - (81 k M^2 - 16) \arctan \left( \frac{9 k M}{4}\right) \right]}\, \cdot\\
		& \left[ 256 \left(729 M^2 (m_1-n_1)+104\right) \right.\\
		&  +405 M^2 p (1323 \ln 3-1276)\\
		& \left. +432 M^2 q (28+621 \ln 3) \right],\\
		c_2 =&  \frac{4 k^{5/2} M \left(-68 + 729 M^2 (4 - 5 \ln 3)q\right)}{9 k M (-16 + 27 k M^2) + 64 \arctan \left( \frac{9 k M}{4}\right)}.
	\end{split}
\end{equation}
At this point, there is a large amount of mathematical freedom in the system, and to proceed we need to impose physical constraints. Our goal is to to uniquely fix the exterior spacetime, parametrized by $p$ and $q$, and thus predict the value of, e.g., the quadrupole moment of the rotating shell. To do this we need to find a way to fix $c_1$ and $c_2$. There is a simple possibility that singles itself out: to simply set $c_1=c_2=0$ so that the interior metric is pure undeformed AdS. The argument for such a choice would be that the piece of fresh space time that is nucleated inside of the bubble, should not depend on what sits outside of it. The junction conditions are then fully taken care of by the matter components on top of the shell together with the metric on the outside. One should note that from the point of view of the inside AdS metric, the outside asymptotic universe will be viewed as rotating, with the AdS glued to the frame dragged spacetime just outside of the shell. We find that $c_2=0$ fixes $q$, while $c_1=0$ then gives a relation between $p$ and $m_1-n_1$.\footnote{In \cite{Danielsson:2017pvl} only $c_1$ was, implicitly, put to zero while the freedom in $c_2$ was not observed. The case studied there had a $c_2$ different from zero.} 

To get further constraints, we then use the second junction condition to compute the energy momentum tensor $S_{ij}$ on the shell. Following the physical motivation behind the construction of the black shells, this stress tensor should be made of various components. As long as each component is a perfect fluid (with corresponding equations of state), and they share the same velocity vector $u^i$, the total stress tensor will also be of the form of a perfect fluid:
\begin{equation}
	S^{ij} = \left(ρ_\textrm{total}+p_\textrm{total}\right) u^i u^j + p_\textrm{total}\, g^{ij}.
\end{equation}
This places a further restriction on the constants and determines $p,q,(m_1-n_1)$ uniquely. The results obtained are exact to all orders in $k$, but too large to write down. So we present only the leading order result in $1/\left(kM\right)$ for some of the quantities below.
\begin{equation}\label{eq:pqmn}
	\begin{split}
		&p = \frac{64 \left(1557 \ln 3 - 1028\right)}{10935(5 \ln 3 - 4) D} \frac{1}{M²}\\
		&\quad + \frac{32768\left( 135 \ln 3 - 172 \right)}{32805 D^2 }\frac{1}{k M^3} +\frac{1}{M²}\orderMksecond,\\
		&q = -\frac{68}{729 (5 \ln 3 - 4)}\frac{1}{M²},\\
		&\hspace{-6pt}\left( m_1 - n_1 \right) =
		 \frac{2 \left( 316144 -726744 \ln 3 +410103 (\ln 3)² \right)}{6561 (5 \ln 3 - 4) D}\frac{1}{M^2}\\
		&\quad -\frac{128(135 \ln 3 - 172) (1323 \ln 3 - 1276)}{59049 D^2}\frac{1}{k M^3}\\
		& \quad +\frac{1}{M²}\orderMksecond,
	\end{split}
\end{equation}
where we have defined $D=196-261 \ln 3$, which will reappear throughout.
The normalized velocity vector $u^i = \left( 3 + γ a², 0, a β \right)$ is: 
\begin{equation}
	\begin{split}
		γ =& -\frac{4 D \left(729 M^2 n_1+64\right)+128 (135 \ln 3 -172) \sin ²θ}{243 M^2 D}\frac{1}{M²}\\
		& +\frac{256\left( 19344-76136 \ln 3 + 63153 (\ln 3)²\right)}{729D²}\frac{1}{k M³}\\
		&+ \frac{1}{M²}\orderMksecond,\\
		β =& -\frac{64\left( 8+27 M k \right)}{243M² \left( 4 + 9 M k \right)}.
	\end{split}
\end{equation}
The values of $p$ and $q$ numerically evaluate to
\begin{equation}
	p = -\frac{0.295}{M²} - \frac{0.00287}{k M³} + \frac{1}{M²}\orderMksecond,\,  
	q = -\frac{0.625}{M²},
\end{equation}
which determines the quadrupole moment, via eq \eqref{eq:quadrupole}, to be
\begin{equation}
\begin{split}
		\mathscr{M}_2 &= -a²M-\frac{2}{15}a²M³\left(16q-15p\right)\\
		& = -0.926 a² M - \frac{0.00575 a²}{k} + a²M \orderMksecond
\end{split}
\end{equation}
Thus, for this physically natural scenario, the expected quadrupole moment of the black shell is about $7\%$ less than the one of a Kerr black hole.\footnote{This is different from the result in \cite{Danielsson:2017pvl}, which corresponded to another possible solution.}

So far, all results are exact, and we have not made any expansion in $k$. To verify that the shell is made of three components with equations of state of tension, gas, and stiff matter we need to expand in large $k$. This allows us to split the total $S_{ij}$ into its constituents:
\begin{equation}\label{eq:stressenergysplit}
	\begin{split}
		\left(S_{\textrm{brane}}\right)^i _j &= \left(-\frac{k}{4π} + \frac{2}{27Mπ} -\frac{1}{81kM²π} - a²\mathscr{Z}_1\right)δ^i _j,\\
		\left(S_{\textrm{gas}}\right)^i _j   &= \left(\frac{1}{18Mπ} + a² \mathscr{Z}_2\right)u^i u_j\\
		&\, +
		\left(\frac{1}{54Mπ} + \frac{a²}{3} \mathscr{Z}_2\right) δ^i _j,\\
		\left(S_{\textrm{stiff}}\right)^i _j &= \left(\frac{2}{81kM²π} + a² \mathscr{Z}_3 \right)u^i u_j\\
		&\, +\left(\frac{1}{81kM²π}+ \frac{a²}{2} \mathscr{Z}_3\right)δ^i _j,
	\end{split}
\end{equation}
where the factors $\mathscr{Z}_i$ are:
\begin{equation}
	\begin{split}
		\mathscr{Z}_1 =& -\frac{A_1 + 16 A_2 \cos 2θ}{6561 π D}\frac{1}{M³} + \frac{A_3 + 96 A_4 \cos 2θ}{177147 π D}\frac{1}{k M^4}\\
		& + \frac{1}{M³} \orderMksecond,\\
		\mathscr{Z}_2 =& \frac{B_1 D + 32 B_2 \cos 2θ}{4374 π D}\frac{1}{M³},\\
		\mathscr{Z}_3 =& -\frac{4 C_1 D + 192 C_2 \cos 2θ}{59049 π D²}\frac{1}{kM^4} + \frac{1}{M³} \orderMksecond,
	\end{split}
\end{equation}
with
\begin{equation}\label{eq:ZAfactors}
	\begin{split}
		A_1 =&\, 36032-54000 \ln 3 +1458 n_1 D M^2 ,\\
		A_2 =&\, 292 - 765 \ln 3,\\
		A_3 =&\, 4374 n_1 D^2 M^2\\
			& +256 (133648+9 \ln 3  (17361 \ln 3 -31336)),\\
		A_4 =&\, 879536 - 2046648 \ln 3 + 1248291 (\ln 3)²,\\
		B_1 =&\, 352 + 2187 n_1 M^2,\\
		B_2 =&\, 52 + 495 \ln 3,\\
		C_1 =&\, 256 (99 \ln 3 -92) + 729 n_1 D M^2,\\
		C_2 =&\, 299440 - 625080 \ln 3  + 336771 (\ln 3)².
	\end{split}
\end{equation}
In the next section we will examine the energy momentum tensor and its properties more closely.

\section{How the rotating shell supports itself}\label{sec:source}

\noindent
One can verify that the induced energy momentum tensor is covariantly conserved with respect to the induced metric, i.e., $∇_i S^i{}_j = 0$. This is not automatically guaranteed. In general, the covariant derivative of the extrinsic curvature is given by a projection of the bulk Einstein tensor on the shell, using the Gauss-Codazzi equations \cite[eq (3.42)]{Poisson:2009pwt}
\begin{equation}
	∇_i K^i{}_j - ∂_j K = G_{μν}e^μ{}_j n^ν,
\end{equation}
where $G_{μν}$ is the four dimensional Einstein tensor in the bulk, $e^μ{}_j$ is the tangent vector, and $n^ν$ is the vector normal to the shell. In our case this vanishes trivially on the outside since $G_{μν}=0$. It also vanishes on the inside due to the projection. Vanishing of the covariant derivative implies that the combined system can sustain itself such that the total pressure balances the gravitational (and centrifugal) forces without the need for any external force acting from the bulk. However, the covariant derivative of each component of the stress tensor is non-vanishing and they cancel against each other non-trivially. Up to $\orderMksecond$, they read:
\begin{equation}\label{eq:S_brane_gas_explicit}
	\begin{split}
		&∇_i \left(S_\textrm{brane}\right)^i{}_θ = \frac{32 (292 - 765 \ln 3) a² \sin 2θ}{6561 \pi D M^3}\\
		&\quad -\frac{64 (879536+3 \ln 3  (416097 \ln 3 -682216)) a² \sin 2θ}{59049 \pi D^2  k M^4},\\
		&∇_i \left(S_\textrm{gas}\right)^i{}_θ = -\frac{32 (292 - 765 \ln 3) a² \sin 2θ}{6561 \pi D M^3}\\
		&\quad + \frac{128 (19344+\ln 3  (63153 \ln 3 -76136)) a² \sin 2θ}{19683 \pi D^2 k M^4 },\\
		&∇_i \left(S_\textrm{stiff}\right)^i{}_θ =\\
		&\quad \frac{64 (763472+9 \ln 3  (96597 \ln 3 -176648)) a² \sin 2θ}{59049 \pi D^2  k M^4},
	\end{split}
\end{equation}
where $D=196 - 261 \ln 3$, as defined below eq \eqref{eq:pqmn}, and all other components vanish.
As promised, one can note how the nonzero terms add up to zero when summed. The physical interpretation of this is that the individual terms exert mutual forces on each other such that the system is stable. From where do these source terms come? It is natural to assume that they have the same origin as the source terms needed to achieve stability in the case of the non-rotating black shell. More precisely, we argue that eq \eqref{eq:flux_orig} should be generalized to
\be\label{eq:j}
j_i \coloneqq 3\rho_g \left(\alpha\ \frac{\nabla_i a_{_R}}{a_{_R}} + \beta\ \frac{F_i}{2} \right).
\ee
In \cite{Danielsson:2021ykm}, $F$ was determined in terms of the area. A convenient alternative would be to express it in terms of the three dimensional Ricci scalar. In the non-rotating case this is simply given by
\be
R^{(3)}=\frac{2+2\dot{R}^2+4R\ddot{R}}{R^2} \approx \frac{2}{R^2}. 
\ee
Hence, we can, to lowest order in any perturbation, use
\be \label{eq:F}
F_i=-\frac{\nabla_i R^{(3)}}{R^{(3)}}.
\ee
We can now evaluate the source term, as given above in eq \eqref{eq:j} and demand that it equals the flux from the non-zero covariant derivative of the brane/gas in eq \eqref{eq:S_brane_gas_explicit}. At leading order in $\left(1/kM\right)$ this gives,
\begin{equation}
	\begin{split}
		j_θ =& -\frac{1}{9Mπ}\frac{a² \sin 2θ}{M²}\, \cdot\\
		& \left( α \frac{40 (100+243 \ln 3 )}{243 D} +β \frac{2560 (21 \ln 3 -4)}{243 D} \right)\\
		=& \frac{40\alpha  (100+243 \ln 3 )+1280 \beta  (21 \ln 3 -4)}{2187 \pi D}  \frac{a² \sin 2θ}{M³}\\
		 \overset{!}{=}& \frac{32 A_2}{6561 \pi  D}\frac{a² \sin 2θ}{M³},
	\end{split}
\end{equation}
where $A_2=292 - 765 \ln 3$, and $D=196-261 \ln 3$, as defined in sec \ref{sec:review}.
This gives the following constraint on $\alpha$ and $\beta$:
\begin{equation}
	4 (292 - 375 \alpha + 480 \beta) - 45 (68 + 81 \alpha +224 \beta) \ln 3   \overset{!}{=} 0.
\end{equation}
If this relation is satisfied, then the shell can support itself at a critical point. If we now use the Buchdahl constraint $\alpha = \beta +2/3$ obtained in \cite{Danielsson:2021ykm}, we can uniquely solve for $\alpha$ and $\beta$. The result is remarkably simple:
\begin{equation}
	\begin{split}
		\alpha &= \frac{4}{15},\\
		\beta &= -\frac{2}{5}.
	\end{split}
\end{equation}
What is even more remarkable is that this satisfies the constraint $\alpha < 4/9$ and exactly saturates the constraint $6\alpha - \beta>2$. So a rotating bubble at a critical point is just barely compatible with the stability of a non-rotating bubble that accretes such that it always sits at the Buchdahl radius. This is a highly non-trivial result. It is worth noting that this solution for $α,β$ does not depend on the value of $\left( m_1 - n_1 \right)$ that was chosen in order to pick the physically most interesting solution with $c_1 = c_2 = 0$. It appears to be a universal feature of rotating black shells that are made of the same ingredients allowing the total energy momentum tensor to be split in same way as in eq \eqref{eq:stressenergysplit}.

To summarize, the source term compatible with stability of the non-rotating case, as well as the existence of a critical point in the case of a rotating black shell is given by
\be\label{eq:source_final}
j_i \equiv \frac{\rho_g}{5} \left(4\frac{\nabla_i a_{_R}}{a_{_R}} + 3\frac{\nabla_i R^{(3)}}{R^{(3)}} \right).
\ee

\section{Conclusions}\label{sec:conclusions}

\noindent
In summary, we have re-examined the rotating black shell alternatives for black holes proposed in \cite{Danielsson:2017pvl}. A black shell is made of a higher dimensional brane from string theory, wrapping internal cycles and yielding a spherical membrane in four dimensional space time. The membrane carries magnetic fluxes, whose massless vibrations within the world volume of the membrane give rise to a gas of radiation. This is heated up to the local Unruh temperature of the shell, thus equilibrating the system and making it lie at a critical point of the gravitational potential. Recently, \cite{Danielsson:2021ykm} analyzed the stability of the non-rotating black shells of \cite{Danielsson:2017riq} and examined a two parameter family of fluxes to ensure that this critical point is stable. It was concluded that stability against radial perturbations, with damping, requires $α<4/9$ and $6α-β>2$. Furthermore, under an impulsive accretion event, for the shell to settle back to its new Buchdahl radius, it is necessary to have $α=β+2/3$.

In this paper, we have extrapolated the stability criteria of \cite{Danielsson:2021ykm} to slowly rotating black shells, and re-examined the flux exchanged between the brane and gas in this case. We found that the flux indeed supports the shell at a critical point in the gravitational potential. Additionally, requiring that the parameters are such that a non-rotating shell relaxes back to its new Buchdahl radius under accretion, fixes the flux parameters to unique values. Remarkably, despite the complicated numerical factors involved, this yields an extremely simple result: $α=4/15$, $β=-2/5$. These satisfy one of the constraints: $α<4/9$, and exactly saturate the other: $6α-β>2$. So the existence of rotating black shells forces the parameters to saturate the stability criteria obtained in \cite{Danielsson:2021ykm} for non-rotating shells.

This opens up several interesting avenues for further study. First, one needs to make sure that rotating black shells accrete matter in a way consistent with that proposed in \cite{Danielsson:2021ykm} for the non-rotating black shells.
Saturating the constraint $6α-β=2$ in eq \eqref{eq:pert} implies that there is a zero mode of the perturbation. However, since eq \eqref{eq:pert} is valid in the large mass limit $m \gg 1/k$, a careful analysis of the sub-leading terms is needed.
This, along with extending the analysis of \cite{Danielsson:2021ykm} to rotating black shells, will require care and careful consideration when approaching this margin of stability.
Second, given these remarkably simple and promising values, it is of fundamental importance to find an argument based on first principles, for the source term as given in eq \eqref{eq:source_final}. 

Third, the analysis, including the suggested reduction of the quadrupole moment by $7\%$ compared to Kerr, is only valid for slowly spinning black shells. It is of great phenomenological interest to extend this analysis to black shells where the rotation is not slow. Given that most supermassive black holes, for example the one at the center of the Milky way in Sagittarius A*, are expected to be spinning close to maximal \cite{Elvis:2001bn}, such an analysis would be important for astrophysical tests that could become available in the near future.

\begin{acknowledgments}
\noindent
UD wants to thank Luis Lehner and Frans Pretorius for discussions. SG is partially supported by the INFN and the MIUR-PRIN contract 2017CC72MK003.
\end{acknowledgments}

\bibliography{prd_submission}

\begin{thebibliography}{25}%
\makeatletter
\providecommand \@ifxundefined [1]{%
 \@ifx{#1\undefined}
}%
\providecommand \@ifnum [1]{%
 \ifnum #1\expandafter \@firstoftwo
 \else \expandafter \@secondoftwo
 \fi
}%
\providecommand \@ifx [1]{%
 \ifx #1\expandafter \@firstoftwo
 \else \expandafter \@secondoftwo
 \fi
}%
\providecommand \natexlab [1]{#1}%
\providecommand \enquote  [1]{``#1''}%
\providecommand \bibnamefont  [1]{#1}%
\providecommand \bibfnamefont [1]{#1}%
\providecommand \citenamefont [1]{#1}%
\providecommand \href@noop [0]{\@secondoftwo}%
\providecommand \href [0]{\begingroup \@sanitize@url \@href}%
\providecommand \@href[1]{\@@startlink{#1}\@@href}%
\providecommand \@@href[1]{\endgroup#1\@@endlink}%
\providecommand \@sanitize@url [0]{\catcode `\\12\catcode `\$12\catcode
  `\&12\catcode `\#12\catcode `\^12\catcode `\_12\catcode `\%12\relax}%
\providecommand \@@startlink[1]{}%
\providecommand \@@endlink[0]{}%
\providecommand \url  [0]{\begingroup\@sanitize@url \@url }%
\providecommand \@url [1]{\endgroup\@href {#1}{\urlprefix }}%
\providecommand \urlprefix  [0]{URL }%
\providecommand \Eprint [0]{\href }%
\providecommand \doibase [0]{https://doi.org/}%
\providecommand \selectlanguage [0]{\@gobble}%
\providecommand \bibinfo  [0]{\@secondoftwo}%
\providecommand \bibfield  [0]{\@secondoftwo}%
\providecommand \translation [1]{[#1]}%
\providecommand \BibitemOpen [0]{}%
\providecommand \bibitemStop [0]{}%
\providecommand \bibitemNoStop [0]{.\EOS\space}%
\providecommand \EOS [0]{\spacefactor3000\relax}%
\providecommand \BibitemShut  [1]{\csname bibitem#1\endcsname}%
\let\auto@bib@innerbib\@empty
\bibitem [{\citenamefont {Hawking}(1976{\natexlab{a}})}]{Hawking:1976ra}%
  \BibitemOpen
  \bibfield  {author} {\bibinfo {author} {\bibfnamefont {S.~W.}\ \bibnamefont
  {Hawking}},\ }\bibfield  {title} {\bibinfo {title} {{Breakdown of
  Predictability in Gravitational Collapse}},\ }\href
  {https://doi.org/10.1103/PhysRevD.14.2460} {\bibfield  {journal} {\bibinfo
  {journal} {Phys. Rev. D}\ }\textbf {\bibinfo {volume} {14}},\ \bibinfo
  {pages} {2460} (\bibinfo {year} {1976}{\natexlab{a}})}\BibitemShut {NoStop}%
\bibitem [{\citenamefont {{Polchinski}}(2017)}]{2017tasi.conf..353P}%
  \BibitemOpen
  \bibfield  {author} {\bibinfo {author} {\bibfnamefont {J.}~\bibnamefont
  {{Polchinski}}},\ }\bibfield  {title} {\bibinfo {title} {{The Black Hole
  Information Problem}},\ }in\ \href
  {https://doi.org/10.1142/9789813149441\_0006} {\emph {\bibinfo {booktitle}
  {New Frontiers in Fields and Strings (TASI 2015}}},\ \bibinfo {editor}
  {edited by\ \bibinfo {editor} {\bibfnamefont {J.}~\bibnamefont
  {{Polchinski}}}\ and\ \bibinfo {editor} {\bibnamefont {{et al.}}}}\ (\bibinfo
  {year} {2017})\ pp.\ \bibinfo {pages} {353--397},\ \Eprint
  {https://arxiv.org/abs/1609.04036} {arXiv:1609.04036 [hep-th]} \BibitemShut
  {NoStop}%
\bibitem [{\citenamefont {Hawking}(1976{\natexlab{b}})}]{Hawking:1976de}%
  \BibitemOpen
  \bibfield  {author} {\bibinfo {author} {\bibfnamefont {S.~W.}\ \bibnamefont
  {Hawking}},\ }\bibfield  {title} {\bibinfo {title} {{Black Holes and
  Thermodynamics}},\ }\href {https://doi.org/10.1103/PhysRevD.13.191}
  {\bibfield  {journal} {\bibinfo  {journal} {Phys. Rev. D}\ }\textbf {\bibinfo
  {volume} {13}},\ \bibinfo {pages} {191} (\bibinfo {year}
  {1976}{\natexlab{b}})}\BibitemShut {NoStop}%
\bibitem [{\citenamefont {Ruffini}\ and\ \citenamefont
  {Wheeler}(1971)}]{Ruffini:1971bza}%
  \BibitemOpen
  \bibfield  {author} {\bibinfo {author} {\bibfnamefont {R.}~\bibnamefont
  {Ruffini}}\ and\ \bibinfo {author} {\bibfnamefont {J.~A.}\ \bibnamefont
  {Wheeler}},\ }\bibfield  {title} {\bibinfo {title} {{Introducing the black
  hole}},\ }\href {https://doi.org/10.1063/1.3022513} {\bibfield  {journal}
  {\bibinfo  {journal} {Phys. Today}\ }\textbf {\bibinfo {volume} {24}},\
  \bibinfo {pages} {30} (\bibinfo {year} {1971})}\BibitemShut {NoStop}%
\bibitem [{\citenamefont {Strominger}\ and\ \citenamefont
  {Vafa}(1996)}]{Strominger:1996sh}%
  \BibitemOpen
  \bibfield  {author} {\bibinfo {author} {\bibfnamefont {A.}~\bibnamefont
  {Strominger}}\ and\ \bibinfo {author} {\bibfnamefont {C.}~\bibnamefont
  {Vafa}},\ }\bibfield  {title} {\bibinfo {title} {{Microscopic origin of the
  Bekenstein-Hawking entropy}},\ }\href
  {https://doi.org/10.1016/0370-2693(96)00345-0} {\bibfield  {journal}
  {\bibinfo  {journal} {Phys. Lett. B}\ }\textbf {\bibinfo {volume} {379}},\
  \bibinfo {pages} {99} (\bibinfo {year} {1996})},\ \Eprint
  {https://arxiv.org/abs/hep-th/9601029} {arXiv:hep-th/9601029} \BibitemShut
  {NoStop}%
\bibitem [{\citenamefont {Susskind}\ \emph {et~al.}(1993)\citenamefont
  {Susskind}, \citenamefont {Thorlacius},\ and\ \citenamefont
  {Uglum}}]{Susskind:1993if}%
  \BibitemOpen
  \bibfield  {author} {\bibinfo {author} {\bibfnamefont {L.}~\bibnamefont
  {Susskind}}, \bibinfo {author} {\bibfnamefont {L.}~\bibnamefont
  {Thorlacius}},\ and\ \bibinfo {author} {\bibfnamefont {J.}~\bibnamefont
  {Uglum}},\ }\bibfield  {title} {\bibinfo {title} {{The Stretched horizon and
  black hole complementarity}},\ }\href
  {https://doi.org/10.1103/PhysRevD.48.3743} {\bibfield  {journal} {\bibinfo
  {journal} {Phys. Rev. D}\ }\textbf {\bibinfo {volume} {48}},\ \bibinfo
  {pages} {3743} (\bibinfo {year} {1993})},\ \Eprint
  {https://arxiv.org/abs/hep-th/9306069} {arXiv:hep-th/9306069} \BibitemShut
  {NoStop}%
\bibitem [{\citenamefont {Almheiri}\ \emph {et~al.}(2013)\citenamefont
  {Almheiri}, \citenamefont {Marolf}, \citenamefont {Polchinski},\ and\
  \citenamefont {Sully}}]{Almheiri:2012rt}%
  \BibitemOpen
  \bibfield  {author} {\bibinfo {author} {\bibfnamefont {A.}~\bibnamefont
  {Almheiri}}, \bibinfo {author} {\bibfnamefont {D.}~\bibnamefont {Marolf}},
  \bibinfo {author} {\bibfnamefont {J.}~\bibnamefont {Polchinski}},\ and\
  \bibinfo {author} {\bibfnamefont {J.}~\bibnamefont {Sully}},\ }\bibfield
  {title} {\bibinfo {title} {{Black Holes: Complementarity or Firewalls?}},\
  }\href {https://doi.org/10.1007/JHEP02(2013)062} {\bibfield  {journal}
  {\bibinfo  {journal} {JHEP}\ }\textbf {\bibinfo {volume} {02}},\ \bibinfo
  {pages} {062}},\ \Eprint {https://arxiv.org/abs/1207.3123} {arXiv:1207.3123
  [hep-th]} \BibitemShut {NoStop}%
\bibitem [{\citenamefont {Mathur}(2009)}]{Mathur:2009hf}%
  \BibitemOpen
  \bibfield  {author} {\bibinfo {author} {\bibfnamefont {S.~D.}\ \bibnamefont
  {Mathur}},\ }\bibfield  {title} {\bibinfo {title} {{The Information paradox:
  A Pedagogical introduction}},\ }\href
  {https://doi.org/10.1088/0264-9381/26/22/224001} {\bibfield  {journal}
  {\bibinfo  {journal} {Class. Quant. Grav.}\ }\textbf {\bibinfo {volume}
  {26}},\ \bibinfo {pages} {224001} (\bibinfo {year} {2009})},\ \Eprint
  {https://arxiv.org/abs/0909.1038} {arXiv:0909.1038 [hep-th]} \BibitemShut
  {NoStop}%
\bibitem [{\citenamefont {'t~Hooft}(1991)}]{tHooft:1991uqr}%
  \BibitemOpen
  \bibfield  {author} {\bibinfo {author} {\bibfnamefont {G.}~\bibnamefont
  {'t~Hooft}},\ }\bibfield  {title} {\bibinfo {title} {{The Black hole horizon
  as a quantum surface}},\ }\href
  {https://doi.org/10.1088/0031-8949/1991/T36/026} {\bibfield  {journal}
  {\bibinfo  {journal} {Phys. Scripta T}\ }\textbf {\bibinfo {volume} {36}},\
  \bibinfo {pages} {247} (\bibinfo {year} {1991})}\BibitemShut {NoStop}%
\bibitem [{\citenamefont {Lunin}\ and\ \citenamefont
  {Mathur}(2002)}]{Lunin:2002qf}%
  \BibitemOpen
  \bibfield  {author} {\bibinfo {author} {\bibfnamefont {O.}~\bibnamefont
  {Lunin}}\ and\ \bibinfo {author} {\bibfnamefont {S.~D.}\ \bibnamefont
  {Mathur}},\ }\bibfield  {title} {\bibinfo {title} {{Statistical
  interpretation of Bekenstein entropy for systems with a stretched horizon}},\
  }\href {https://doi.org/10.1103/PhysRevLett.88.211303} {\bibfield  {journal}
  {\bibinfo  {journal} {Phys. Rev. Lett.}\ }\textbf {\bibinfo {volume} {88}},\
  \bibinfo {pages} {211303} (\bibinfo {year} {2002})},\ \Eprint
  {https://arxiv.org/abs/hep-th/0202072} {arXiv:hep-th/0202072} \BibitemShut
  {NoStop}%
\bibitem [{\citenamefont {Giusto}\ \emph {et~al.}(2004)\citenamefont {Giusto},
  \citenamefont {Mathur},\ and\ \citenamefont {Saxena}}]{Giusto:2004id}%
  \BibitemOpen
  \bibfield  {author} {\bibinfo {author} {\bibfnamefont {S.}~\bibnamefont
  {Giusto}}, \bibinfo {author} {\bibfnamefont {S.~D.}\ \bibnamefont {Mathur}},\
  and\ \bibinfo {author} {\bibfnamefont {A.}~\bibnamefont {Saxena}},\
  }\bibfield  {title} {\bibinfo {title} {{Dual geometries for a set of 3-charge
  microstates}},\ }\href {https://doi.org/10.1016/j.nuclphysb.2004.09.001}
  {\bibfield  {journal} {\bibinfo  {journal} {Nucl. Phys. B}\ }\textbf
  {\bibinfo {volume} {701}},\ \bibinfo {pages} {357} (\bibinfo {year}
  {2004})},\ \Eprint {https://arxiv.org/abs/hep-th/0405017}
  {arXiv:hep-th/0405017} \BibitemShut {NoStop}%
\bibitem [{\citenamefont {Cardoso}\ and\ \citenamefont
  {Pani}(2019)}]{Cardoso:2019rvt}%
  \BibitemOpen
  \bibfield  {author} {\bibinfo {author} {\bibfnamefont {V.}~\bibnamefont
  {Cardoso}}\ and\ \bibinfo {author} {\bibfnamefont {P.}~\bibnamefont {Pani}},\
  }\bibfield  {title} {\bibinfo {title} {{Testing the nature of dark compact
  objects: a status report}},\ }\href
  {https://doi.org/10.1007/s41114-019-0020-4} {\bibfield  {journal} {\bibinfo
  {journal} {Living Rev. Rel.}\ }\textbf {\bibinfo {volume} {22}},\ \bibinfo
  {pages} {4} (\bibinfo {year} {2019})},\ \Eprint
  {https://arxiv.org/abs/1904.05363} {arXiv:1904.05363 [gr-qc]} \BibitemShut
  {NoStop}%
\bibitem [{\citenamefont {Danielsson}\ \emph {et~al.}(2017)\citenamefont
  {Danielsson}, \citenamefont {Dibitetto},\ and\ \citenamefont
  {Giri}}]{Danielsson:2017riq}%
  \BibitemOpen
  \bibfield  {author} {\bibinfo {author} {\bibfnamefont {U.~H.}\ \bibnamefont
  {Danielsson}}, \bibinfo {author} {\bibfnamefont {G.}~\bibnamefont
  {Dibitetto}},\ and\ \bibinfo {author} {\bibfnamefont {S.}~\bibnamefont
  {Giri}},\ }\bibfield  {title} {\bibinfo {title} {{Black holes as bubbles of
  AdS}},\ }\href {https://doi.org/10.1007/JHEP10(2017)171} {\bibfield
  {journal} {\bibinfo  {journal} {JHEP}\ }\textbf {\bibinfo {volume} {10}},\
  \bibinfo {pages} {171}},\ \Eprint {https://arxiv.org/abs/1705.10172}
  {arXiv:1705.10172 [hep-th]} \BibitemShut {NoStop}%
\bibitem [{\citenamefont {Danielsson}\ and\ \citenamefont
  {Giri}(2018)}]{Danielsson:2017pvl}%
  \BibitemOpen
  \bibfield  {author} {\bibinfo {author} {\bibfnamefont {U.}~\bibnamefont
  {Danielsson}}\ and\ \bibinfo {author} {\bibfnamefont {S.}~\bibnamefont
  {Giri}},\ }\bibfield  {title} {\bibinfo {title} {{Observational signatures
  from horizonless black shells imitating rotating black holes}},\ }\href
  {https://doi.org/10.1007/JHEP07(2018)070} {\bibfield  {journal} {\bibinfo
  {journal} {JHEP}\ }\textbf {\bibinfo {volume} {07}},\ \bibinfo {pages}
  {070}},\ \Eprint {https://arxiv.org/abs/1712.00511} {arXiv:1712.00511
  [hep-th]} \BibitemShut {NoStop}%
\bibitem [{\citenamefont {Danielsson}\ \emph {et~al.}(2021)\citenamefont
  {Danielsson}, \citenamefont {Lehner},\ and\ \citenamefont
  {Pretorius}}]{Danielsson:2021ykm}%
  \BibitemOpen
  \bibfield  {author} {\bibinfo {author} {\bibfnamefont {U.}~\bibnamefont
  {Danielsson}}, \bibinfo {author} {\bibfnamefont {L.}~\bibnamefont {Lehner}},\
  and\ \bibinfo {author} {\bibfnamefont {F.}~\bibnamefont {Pretorius}},\
  }\bibfield  {title} {\bibinfo {title} {{Dynamics and Observational Signatures
  of Shell-like Black Hole Mimickers}},\ }\href@noop {} {\  (\bibinfo {year}
  {2021})},\ \Eprint {https://arxiv.org/abs/2109.09814} {arXiv:2109.09814
  [gr-qc]} \BibitemShut {NoStop}%
\bibitem [{\citenamefont {Coleman}\ and\ \citenamefont
  {De~Luccia}(1980)}]{Coleman:1980aw}%
  \BibitemOpen
  \bibfield  {author} {\bibinfo {author} {\bibfnamefont {S.~R.}\ \bibnamefont
  {Coleman}}\ and\ \bibinfo {author} {\bibfnamefont {F.}~\bibnamefont
  {De~Luccia}},\ }\bibfield  {title} {\bibinfo {title} {{Gravitational Effects
  on and of Vacuum Decay}},\ }\href {https://doi.org/10.1103/PhysRevD.21.3305}
  {\bibfield  {journal} {\bibinfo  {journal} {Phys. Rev. D}\ }\textbf {\bibinfo
  {volume} {21}},\ \bibinfo {pages} {3305} (\bibinfo {year}
  {1980})}\BibitemShut {NoStop}%
\bibitem [{\citenamefont {Israel}(1966)}]{Israel:1966rt}%
  \BibitemOpen
  \bibfield  {author} {\bibinfo {author} {\bibfnamefont {W.}~\bibnamefont
  {Israel}},\ }\bibfield  {title} {\bibinfo {title} {{Singular hypersurfaces
  and thin shells in general relativity}},\ }\href
  {https://doi.org/10.1007/BF02710419} {\bibfield  {journal} {\bibinfo
  {journal} {Nuovo Cim. B}\ }\textbf {\bibinfo {volume} {44S10}},\ \bibinfo
  {pages} {1} (\bibinfo {year} {1966})},\ \bibinfo {note} {[Erratum: Nuovo
  Cim.B 48, 463 (1967)]}\BibitemShut {NoStop}%
\bibitem [{\citenamefont {Lanczos}()}]{Lanczos:1927}%
  \BibitemOpen
  \bibfield  {author} {\bibinfo {author} {\bibfnamefont {K.}~\bibnamefont
  {Lanczos}},\ }\bibfield  {title} {\bibinfo {title} {Flächenhafte verteilung
  der materie in der einsteinschen gravitationstheorie},\ }\href
  {https://doi.org/https://doi.org/10.1002/andp.19243791403} {\bibfield
  {journal} {\bibinfo  {journal} {Annalen der Physik}\ }\textbf {\bibinfo
  {volume} {379}},\ \bibinfo {pages} {518}}\BibitemShut {NoStop}%
\bibitem [{\citenamefont {Sen}()}]{Sen:1924}%
  \BibitemOpen
  \bibfield  {author} {\bibinfo {author} {\bibfnamefont {N.}~\bibnamefont
  {Sen}},\ }\bibfield  {title} {\bibinfo {title} {Über die grenzbedingungen
  des schwerefeldes an unstetigkeitsflächen},\ }\href
  {https://doi.org/https://doi.org/10.1002/andp.19243780505} {\bibfield
  {journal} {\bibinfo  {journal} {Annalen der Physik}\ }\textbf {\bibinfo
  {volume} {378}},\ \bibinfo {pages} {365}}\BibitemShut {NoStop}%
\bibitem [{\citenamefont {Manko}\ and\ \citenamefont
  {Novikov}(1992)}]{Manko_1992}%
  \BibitemOpen
  \bibfield  {author} {\bibinfo {author} {\bibfnamefont {V.~S.}\ \bibnamefont
  {Manko}}\ and\ \bibinfo {author} {\bibfnamefont {I.~D.}\ \bibnamefont
  {Novikov}},\ }\bibfield  {title} {\bibinfo {title} {Generalizations of the
  kerr and kerr-newman metrics possessing an arbitrary set of mass-multipole
  moments},\ }\href {https://doi.org/10.1088/0264-9381/9/11/013} {\bibfield
  {journal} {\bibinfo  {journal} {Classical and Quantum Gravity}\ }\textbf
  {\bibinfo {volume} {9}},\ \bibinfo {pages} {2477} (\bibinfo {year}
  {1992})}\BibitemShut {NoStop}%
\bibitem [{\citenamefont {Hartle}\ and\ \citenamefont
  {Thorne}(1968)}]{Hartle:1968si}%
  \BibitemOpen
  \bibfield  {author} {\bibinfo {author} {\bibfnamefont {J.~B.}\ \bibnamefont
  {Hartle}}\ and\ \bibinfo {author} {\bibfnamefont {K.~S.}\ \bibnamefont
  {Thorne}},\ }\bibfield  {title} {\bibinfo {title} {{Slowly Rotating
  Relativistic Stars. II. Models for Neutron Stars and Supermassive Stars}},\
  }\href {https://doi.org/10.1086/149707} {\bibfield  {journal} {\bibinfo
  {journal} {Astrophys. J.}\ }\textbf {\bibinfo {volume} {153}},\ \bibinfo
  {pages} {807} (\bibinfo {year} {1968})}\BibitemShut {NoStop}%
\bibitem [{\citenamefont {Geroch}(1970)}]{Geroch:1970cd}%
  \BibitemOpen
  \bibfield  {author} {\bibinfo {author} {\bibfnamefont {R.~P.}\ \bibnamefont
  {Geroch}},\ }\bibfield  {title} {\bibinfo {title} {{Multipole moments. II.
  Curved space}},\ }\href {https://doi.org/10.1063/1.1665427} {\bibfield
  {journal} {\bibinfo  {journal} {J. Math. Phys.}\ }\textbf {\bibinfo {volume}
  {11}},\ \bibinfo {pages} {2580} (\bibinfo {year} {1970})}\BibitemShut
  {NoStop}%
\bibitem [{\citenamefont {Hansen}(1974)}]{Hansen:1974zz}%
  \BibitemOpen
  \bibfield  {author} {\bibinfo {author} {\bibfnamefont {R.~O.}\ \bibnamefont
  {Hansen}},\ }\bibfield  {title} {\bibinfo {title} {{Multipole moments of
  stationary space-times}},\ }\href {https://doi.org/10.1063/1.1666501}
  {\bibfield  {journal} {\bibinfo  {journal} {J. Math. Phys.}\ }\textbf
  {\bibinfo {volume} {15}},\ \bibinfo {pages} {46} (\bibinfo {year}
  {1974})}\BibitemShut {NoStop}%
\bibitem [{\citenamefont {Poisson}()}]{Poisson:2009pwt}%
  \BibitemOpen
  \bibfield  {author} {\bibinfo {author} {\bibfnamefont {E.}~\bibnamefont
  {Poisson}},\ }\bibfield  {title} {\bibinfo {title} {{A Relativist's Toolkit:
  The Mathematics of Black-Hole Mechanics}},\ }\href@noop {} {\ }\BibitemShut
  {NoStop}%
\bibitem [{\citenamefont {Elvis}\ \emph {et~al.}(2002)\citenamefont {Elvis},
  \citenamefont {Risaliti},\ and\ \citenamefont {Zamorani}}]{Elvis:2001bn}%
  \BibitemOpen
  \bibfield  {author} {\bibinfo {author} {\bibfnamefont {M.}~\bibnamefont
  {Elvis}}, \bibinfo {author} {\bibfnamefont {G.}~\bibnamefont {Risaliti}},\
  and\ \bibinfo {author} {\bibfnamefont {G.}~\bibnamefont {Zamorani}},\
  }\bibfield  {title} {\bibinfo {title} {{Most supermassive black holes must be
  rapidly rotating}},\ }\href {https://doi.org/10.1086/339197} {\bibfield
  {journal} {\bibinfo  {journal} {Astrophys. J. Lett.}\ }\textbf {\bibinfo
  {volume} {565}},\ \bibinfo {pages} {L75} (\bibinfo {year} {2002})},\ \Eprint
  {https://arxiv.org/abs/astro-ph/0112413} {arXiv:astro-ph/0112413}
  \BibitemShut {NoStop}%
\end{thebibliography}%

\end{document}